\documentclass[sn-mathphys-num]{sn-jnl}

\usepackage{graphicx}%
\usepackage{multirow}%
\usepackage{amsmath,amssymb,amsfonts}%
\usepackage{amsthm}%
\usepackage{mathrsfs}%
\usepackage[title]{appendix}%
\usepackage{xcolor}%
\usepackage{textcomp}%
\usepackage{manyfoot}%
\usepackage{booktabs}%
\usepackage{algorithm}%
\usepackage{algorithmicx}%
\usepackage{algpseudocode}%
\usepackage{listings}%

\usepackage{setspace}

\usepackage{lmodern}

\usepackage{lineno}

\begin{document}

\title[Calculation of The Abundance of $ ^{187}Re $ -- $ ^{187}Os $ Nuclear Clock Nuclides in S-process and Sensitivity Analysis of Maxwellian-Averaged Neutron Capture Cross Sections]{Calculation of The Abundance of $ ^{187}Re $ -- $ ^{187}Os $ Nuclear Clock Nuclides in S-process and Sensitivity Analysis of Maxwellian-Averaged Neutron Capture Cross Sections}

\author[1]{\fnm{Xinyu} \sur{Dong}}\email{dingzfcpp4141@126.com}

\author[2]{\fnm{Yixuan} \sur{Qiu}}\email{yii\underline{~}xuan@163.com}

\author*[1]{\fnm{Kaisu} \sur{Wu}}\email{wuks@mail.buct.edu.cn}

\affil[1]{\orgdiv{College of Mathematics and Physics}, \orgname{Beijing University of Chemical Technology}, \orgaddress{\street{15 North Third Ring East Road}, \city{Beijing}, \postcode{100029}, \state{Beijing}, \country{China}}}

\affil[2]{\orgdiv{SEU-Monash Joint Graduate School}, \orgname{Southeast University}, \orgaddress{\street{399 Linquan Road}, \city{Suzhou}, \postcode{215000}, \state{Jiangsu}, \country{China}}}

\abstract{In this paper, the network equations calculation of $ ^{187}Re $ -- $ ^{187}Os $ clock-related nuclide abundance in s-process is studied, and the sensitivities of  Maxwellian-Averaged neutron capture cross sections for each nuclide are analyzed in detail. Firstly, basing nuclear physical parameters, we give the branching s-process reaction network from $ ^{184}W $ to $ ^{190}Os $, and establish the corresponding network equations. Using a single path s-process approximation, we obtain an analytical expression of the seed nuclide $ ^{183}W $ abundance of our branching network. Because of the stiffness of the system of network equations, we use the semi-implicit Runge-Kutta method to give the numerical solution of the network equations, and thus obtain the abundance of each nuclide related to the $ ^{187}Re $ -- $ ^{187}Os $ nuclear clock in the s-process. Finally, with the numerical solution, the sensitivity analysis of the Maxwellian-Averaged neutron capture cross sections of the nuclear reaction involved in the $ ^{187}Re $ -- $ ^{187}Os $ nuclear clock network equations is carried out. Therefore, we find that in s-process, the neutron capture reaction $ {}^{184} W+n \rightarrow {}^{185}W $ has the greatest influence on the $ ^{187}Re $ -- $ ^{187}Os $ nuclear clock reaction network, and the neutron capture reaction $ ^{186} W+n \rightarrow {}^{187}W $ has the greatest effect on the particular nuclides $ ^{187}Re $ and $ ^{187}Os $. So the measurements of these two Maxwellian-Averaged neutron capture cross sections deserve the attention of experimental nuclear physicists. }



\maketitle

\section{Introduction} \label{sec1}

One of the key topics of cosmological chronology is determining the age of celestial bodies  \cite{vangioni1990astrophysical,arnould1999nuclear}, which provides fundamental information on the formation and evolution of celestial bodies, and is one of the most important parameters in astrophysics and cosmology. Cosmological chronology proposes that the long-lived radionuclides can serve as cosmic nuclear clocks  \cite{rutherford1929origin}, such as $ ^{40}K $ \cite{burbidge1957synthesis}, $ ^{87}Rb $ \cite{clayton1964cosmoradiogenic,clayton1969isotopic}, $ ^{176}Lu $ \cite{hayakawa2004evidence}, $ ^{187}Re $ \cite{clayton1964cosmoradiogenic,clayton1969isotopic}, $ ^{232}Th $ and $ ^{238}U $ \cite{burbidge1957synthesis,fowler1960nuclear}.

Among them, the abundance ratio of $ ^{187}Re $ and its decay nuclei suggested by Clayton as the cosmic nuclear clock has unique significance. Besides, $ ^{187}Re $ is mainly produced by the r-process, $ ^{186}Os $ and $ ^{187}Os $ are mainly produced by the s-process and part of $ ^{187}Os $ is produced by $ ^{187}Re $ through $\beta$-decay. The ground state of $ ^{187}Re $ decays to $ ^{187}Os $ \cite{bosch1996observation} with a half-life of $ 4.35 \times 10^{10} $ years \cite{lindner1986direct}. Based on the above analysis, it can be seen that the reaction process is a branching s-process, and the production rates of $ ^{186}Os $ and $ ^{187}Os $ are independent of the uncertainty of the r-process. The $ ^{187}Re $ -- $ ^{187}Os $ nuclear clock is also less affected by late perturbation events than the $ ^{87}Rb $ -- $ ^{87}Sr $ nuclear clock. And compared with the $ ^{40}Kr $ -- $ ^{40}Ar $ nuclear clock, the $ ^{187}Re $ -- $ ^{187}Os $ nuclear clock has less nuclear fission caused by cosmic rays \cite{clayton1961neutron}. Therefore, there is less error in determining the stellar age by studying the $ ^{187}Re $ -- $ ^{187}Os $ nuclear clock. 

In 1982, J. M. Luck \cite{luck1980187} pointed out that the half-life of $ ^{187}Re $ was similar to the age of the galaxy in order of magnitude and estimated the age range of the galaxy through the chemical experimental study of meteorite composition. Takahashi \cite{takahashi1983nuclear,yokoi1983re} and others found an uncertainty in the calibration of the nuclear clock. In 2002, Xixiang Bai \cite{bai2003bound} analysed the bound-state $\beta$-decay and its astrophysical significance, and proposed a calibration direction of the  $ ^{187}Re $ -- $ ^{187}Os $ cosmic nuclear clock, but did not carry out the actual data analysis. T. Hayakawa \cite{hayakawa2005new}, considered the effect of isomeric states. 

$ ^{187}Re $ nuclei are mainly obtained from r-process products by $\beta$-decay after the end of the r-process and located outside the main path of the s-process. The s-process nuclei $ ^{186}Os $ and $ ^{187}Os $ are not directly produced by the r-process because they are shielded by stable nuclei $ ^{186}W $ and $ ^{187}Re $. Therefore, the pure s-process nuclei $ ^{186}Os $ can be used to normalize the s-process nuclei abundance in this mass region \cite{hayakawa2005new}. Notably, the $ ^{187}Re $ -- $ ^{187}Os $ nuclear clock has the advantage that it avoids the uncertainty in the initial abundances calculated by the r-process model. Also, since part of $ ^{187}Os $ is produced by the r-process nucleus $ ^{187}Re $ through $\beta$-decay, the nuclear clock can be calibrated by subtracting the $ ^{187}Re $ and $ ^{187}Os $ abundances from the s-process contributions to these nuclei. So it is important to perform detailed calculations of the nuclide abundances of the  $ ^{187}Re $ -- $ ^{187}Os $ nuclear clock in the s-process, which is one of the aims of this study.

To calculate the abundance of $ ^{187}Re $ -- $ ^{187}Os $ nuclear clock network nuclides in s-process in detail, it is necessary to construct the related reaction network, establish the network equations and solve the equations. The coefficients of equations are the Maxwellian-Averaged neutron capture cross sections and decay rates of the each nuclides. These data are currently obtained by nuclear physicists in experiments \cite{fujii2010neutron,kappeler1991s,shizuma2005photodisintegration,segawa2007neutron}. However, nuclear physics experiments usually contain experimental errors, which have an impact on subsequent calculations of the nuclide abundance. Therefore, it is necessary to analyze the sensitivity of Maxwellian-Averaged neutron capture cross sections. In this paper, the calculation of sensitivity is similar to the Ref.~\cite{gao2021network}. By using the control variable method, the reaction cross section of a certain neutron capture reaction is changed from $ -20\% $ to $ 20\% $ each time in the step of $ H=10\% $. After changing Maxwellian-Averaged neutron capture cross section each time, the abundance of all nuclides will inevitably change. Unlike that of the Ref.~\cite{gao2021network}, we define sensitivity as the sum of the absolute value of the change in abundance of each nuclide after changing Maxwell-Average neutron capture cross sections multiplied by the calculated step (which is actually a 1-norm of the function), so that the effect of this neutron capture reaction on the whole network is obtained. Therefore, it can be identified that the neutron capture reaction that has the largest effect on the abundance of $ ^{187}Re $ -- $ ^{187}Os $ nuclear clock-related nuclides in the s-process is $ {}^{184} W+n \rightarrow {}^{185} W $. 
Subsequently, we use sensitivity analysis to obtain that the neutron capture reactions with the greatest impact on the special nuclides $ ^{187}Re $ and $ ^{187}Os $ is $ ^{186}W+n \rightarrow {}^{187}W $. We recommend that nuclear physics experimenters pay attention to these two nuclear reactions. 

This paper is organized as follows. In the second section, according to the nuclear physics data, the s-process branching network path is given and network differential equations are obtained and simplified. We first find the analytical solution of the seed nuclide $ ^{183}W $ in the network, then get the numerical solution of differential equations by using the semi-implicit fourth-order Runge-Kutta method of the stiff equations and obtain the abundance variation curves for each nuclide. In the third section, we carry out the sensitivity analysis of the Maxwellian-Averaged neutron capture cross sections of the nuclear reactions involved in the $ ^{187}Re $ -- $ ^{187}Os $ nuclear clock network equations. The Maxwellian-Averaged neutron capture cross sections vary from $ 80\% $ to $ 120\%  $ and we obtain the Maxwellian-Averaged neutron capture cross sections that have the greatest effect on the total reaction and on the particular nuclides ($ ^{187}Re $ and $ ^{187}Os $). The summary is presented in section 4.

\section{Reaction Network and Network Equations of the $ ^{187}Re $ -- $ ^{187}Os $ Nuclear Clock in the S-process}\label{sec2}
The nuclear reaction network of $ ^{187}Re $ -- $ ^{187}Os $ nuclear clock in the s-process is a branching network. There are unstable nuclides that half-lives of them are same with the neutron capture time scale in the path of the s-process \cite{koloczek2016sensitivity}. When the s-process passes through these nuclides, neutron capture and $\beta$-decay occur simultaneously \cite{hoyle1956origin}, so these nuclides are the branching points of the reaction process. 

The s-process starts with $ ^{56}Fe $. After a series of neutron captures and decays, $ ^{183}W $ is synthesized. Then, $ ^{183}W $ captures neutron to synthesize $ ^{184}W $, and next the nuclide enters the nuclear reaction network of the $ ^{187}Re $ -- $ ^{187}Os $ nuclear clock. We note that the half-life of $ ^{186}Re $ is $ 3.72 $ days, yet the the typical time scale of neutron capture in the s-process is usually $ 10$ -- $100 $ years. So the $\beta$-decay of $ ^{186}Re $ occurs before neutron capture and the reaction $ {}^{186} Re+n \rightarrow^{187} Re $ is ignored. Based on the above analysis, we give the path diagram of the nuclear reaction network of the $ ^{187}Re $ -- $ ^{187}Os $ nuclear clock in the s-process. 

\begin{figure}[ht]
	\centering
	\includegraphics[scale=0.4]{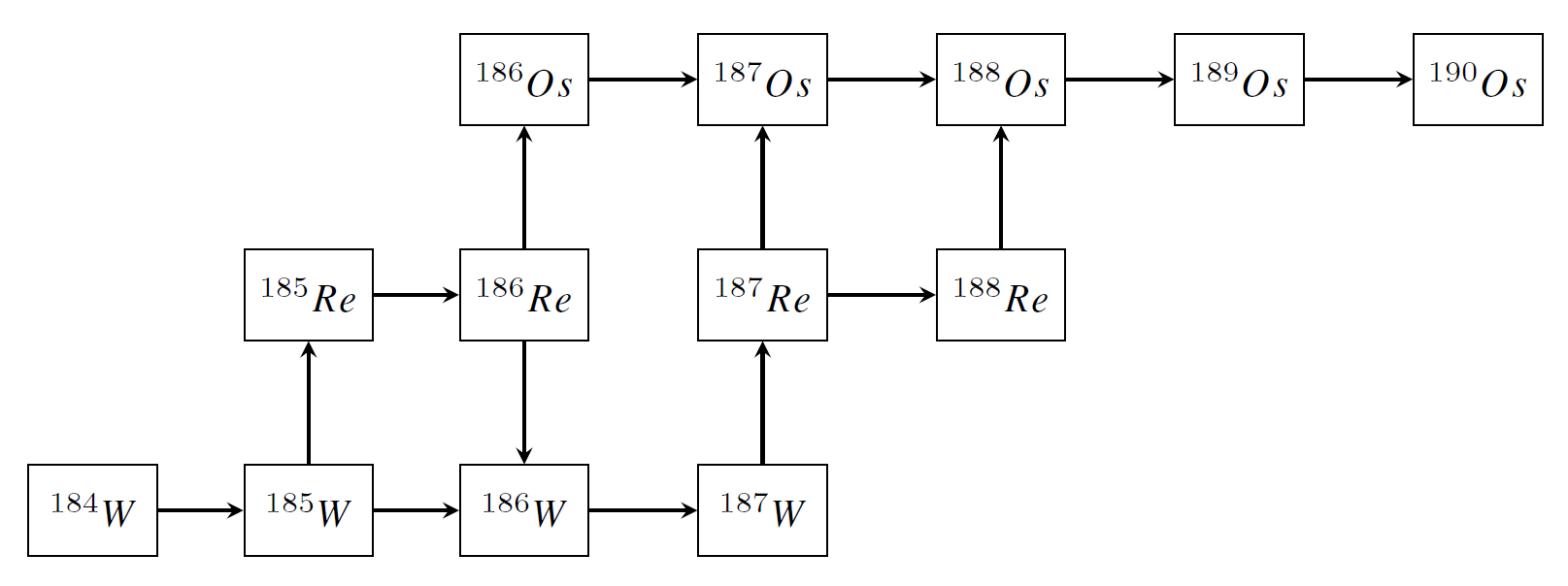}
	\caption{Path diagram of branching s-process  network from $ ^{184}W $ to $ ^{190}Os $. }
	\label{fig:1}
\end{figure}

According to the nuclear reaction network path shown in Figure \ref{fig:1}, the corresponding differential equations of the branching network can be obtained as follows:

\begin{align}
	\frac{\mathrm{d} N( ^{184}W)}{\mathrm{d} t} =&\lambda_{n}(^{183}W)N(^{183}W)-\lambda_{n}(^{184}W)N(^{184}W), \nonumber \\
	\frac{\mathrm{d} N( ^{185}W)}{\mathrm{d} t} =&\lambda_{n}(^{184}W)N(^{184}W)-[\lambda_{\raisebox{0mm}{-}}(^{185}W)+\lambda_{n}(^{185}W)]N(^{185}W),   \nonumber  \\
	\frac{\mathrm{d} N( ^{185}Re)}{\mathrm{d} t} =&\lambda_{\raisebox{0mm}{-}}(^{185}W)N(^{185}W)-\lambda_{n}(^{185}Re)N(^{185}Re),   \nonumber  \\
	\frac{\mathrm{d} N( ^{186}Re)}{\mathrm{d} t} =&\lambda_{n}(^{185}Re)N(^{185}Re)-[\lambda_{\raisebox{0mm}{-}}(^{186}Re)\nonumber+\lambda_{ec}(^{186}Re)]N(^{186}Re),  \nonumber  \\
	\frac{\mathrm{d} N( ^{186}W)}{\mathrm{d} t} =&\lambda_{n}(^{185}W)N(^{185}W)+\lambda_{ec}(^{186}Re)N(^{186}Re)-\lambda_{n}(^{186}W)N(^{186}W), \nonumber   \\
	\frac{\mathrm{d} N( ^{187}W)}{\mathrm{d} t} =&\lambda_{n}(^{186}W)N(^{186}W)-\lambda_{\raisebox{0mm}{-}}(^{187}W)N(^{187}W),  \nonumber  \\
	\frac{\mathrm{d} N( ^{186}Os)}{\mathrm{d} t} =&\lambda_{\raisebox{0mm}{-}}(^{186}Re)N(^{186}Re)-\lambda_{n}(^{186}Os)N(^{186}Os),  \nonumber \\
	\frac{\mathrm{d} N( ^{187}Re)}{\mathrm{d} t} =&\lambda_{\raisebox{0mm}{-}}(^{187}W)N(^{187}W)-[\lambda_{\raisebox{0mm}{-}}(^{187}Re)\nonumber+\lambda_{n}(^{187}Re)]N(^{187}Re),   \nonumber \\
	\frac{\mathrm{d} N( ^{187}Os)}{\mathrm{d} t} 
	=&\lambda_{n}(^{186}Os)N(^{186}Os)+\lambda_{\raisebox{0mm}{-}}(^{187}Re)N(^{187}Re)-\lambda_{n}(^{187}Os)N(^{187}Os),   \nonumber \\
	\frac{\mathrm{d} N( ^{188}Re)}{\mathrm{d} t} =&\lambda_{n}(^{187}Re)N(^{187}Re)-\lambda_{\raisebox{0mm}{-}}(^{188}Re)N(^{188}Re),  \nonumber  \\
	\frac{\mathrm{d} N( ^{188}Os)}{\mathrm{d} t} =&\lambda_{n}(^{187}Os)N(^{187}Os)+\lambda_{\raisebox{0mm}{-}}(^{188}Re)N(^{188}Re)-\lambda_{n}(^{188}Os)N(^{188}Os),  \nonumber  \\
	\frac{\mathrm{d} N( ^{189}Os)}{\mathrm{d} t} =&\lambda_{n}(^{188}Os)N(^{188}Os)-\lambda_{n}(^{189}Os)N(^{189}Os).
\end{align}

Where $ N(A) $ is the abundance of nuclide A; $ \lambda_{n} = n_{n}<\sigma v> $ is the reaction rate of neutron capture; $ n_{n} $ is the number density of neutrons in the reaction process; $ <\sigma v> $ is the probability of the incident neutron reacting with the nucleus; $ \lambda_{\raisebox{0mm}{-}} $ is  beta decay rate;  $ \lambda_{ec} $  is the electron capture rate in $ \varepsilon  $ decay. 

Similar to the Ref.~\cite{pan2020the,ward1976s}, we use the variable substitution introduced by Clayton D.D \cite{clayton1961neutron}. 

\begin{equation} 
	\tau\equiv\int_{0}^{t}n_{n}(t^{'})v_{T}dt^{'}. 
\end{equation}	

Meanwhile, the abundance of each nuclide is normalized to Fe \cite{clayton1961neutron} with the following transformation, 

\begin{equation} 
	\psi(A)=\sigma(A)N(A)/N_{0}(^{56}Fe). 
\end{equation}	

Here $\sigma(A)$ is the Maxwellian-Averaged neutron capture cross section of the nuclide $ A $. Then the abundance of Fe is set to $ 1 $. Nuclides with very short half-lives ($ T<1 $ day) are treated as extremely unstable nuclides, which abundance changes approximately to $ 0 $. The simplified differential equations are arranged as follows. 

\begin{equation} 
	\frac{\mathrm{d}}{\mathrm{d}\tau}\Psi=M\Psi+\sigma(^{184}W)\psi(^{183}W)e_{1}. 
	\label{eq:4}
\end{equation}	

\begin{equation} 
	\Psi\equiv\begin{bmatrix}
		\psi(^{184}W)\\
		\psi(^{185}Re)\\
		\psi(^{186}W)\\
		\psi(^{186}Os)\\
		\psi(^{187}Re)\\
		\psi(^{187}Os)\\
		\psi(^{188}Os)\\
		\psi(^{189}Os)
	\end{bmatrix},e_{1}\equiv\begin{bmatrix}
		1\\
		0\\
		0\\
		0\\
		0\\
		0\\
		0\\
		0
	\end{bmatrix}. 
\end{equation}

\begin{equation}
	M\equiv\begin{bmatrix}
		M_{1}& M_{2}
	\end{bmatrix}. 
	\label{eq:6}
\end{equation}

The submatrices of the block matrix in Eq.~(\ref{eq:6}) are shown below.

\begin{equation}
	M_{1}=\begin{bmatrix}
		\raisebox{0mm}{-}\sigma(^{184}W)&	0&0&			\\
		
		\dfrac{\lambda_{\raisebox{0mm}{-}}(^{185}W)}{n_{n}v_{T}} \dfrac{\sigma(^{185}Re)}{\sigma(^{185}W)+\frac{\lambda_{\raisebox{0mm}{-}}(^{185}W)}{n_{n} v_{T}}} &\raisebox{0mm}{-}\sigma(^{185}Re)&0&\\
		
		\dfrac{\sigma(^{185}W) \sigma(^{186}W)}{\sigma(^{185}W)+\frac{\lambda_{\raisebox{0mm}{-}}(^{185}W)}{n_{n} v_{T}}} &\dfrac{\lambda_{ec}(^{186}Re) \sigma(^{186}W)}{\lambda_{ec}(^{186}Re)+\lambda_{\raisebox{0mm}{-}}(^{186}Re)}&\raisebox{0mm}{-}\sigma(^{186}W)& \\
		
		0&\dfrac{\lambda_{\raisebox{0mm}{-}}(^{186}Re) \sigma(^{186}Os)}{\lambda_{ec}(^{186}Re)+\lambda_{\raisebox{0mm}{-}}(^{186}Re)} &0&\\
		
		0&0&\sigma(^{187}Re) &\\
		
		0& 0&0& \\
		
		0&0&0&\\
		
		0 &0 &0&                              
	\end{bmatrix}. 
\end{equation}

\begin{equation}
	M_{2}=\begin{bmatrix}
		0&0&0&  0& 0\\
		0&0&0&  0& 0\\
		0&0&0&  0 &0\\
		\raisebox{0mm}{-}\sigma(^{186}Os)&0&0&  0& 0\\
		
		0&\raisebox{0mm}{-}[ \sigma(^{187}Re)+\frac{\lambda_{\raisebox{0mm}{-}}(^{187}Re)}{n_{n} v_{T}} ]&0&  0&  0\\
		
		\sigma(^{187}Os)&\frac{\lambda_{\raisebox{0mm}{-}}(^{187}Re)}{n_{n} v_{T}} \frac{\sigma(^{187}Os)}{\sigma(^{187}Re)}&\raisebox{0mm}{-}\sigma(^{187}Os)&  0&  0\\
		
		0&\sigma(^{188}Os) &\sigma(^{188}Os)&  \raisebox{0mm}{-}\sigma(^{188}Os)&  0\\
		
		0&0&0&  \sigma(^{189}Os)&  \raisebox{0mm}{-}\sigma(^{189}Os)
	\end{bmatrix}. 
\end{equation}

From the expression of Eq.~(\ref{eq:4}), it is easy to see that to solve the network equations, the abundance of $ ^{183}W $ seed nuclide of the network must be obtained first. Since the half-life of $ ^{183}W $ is $ 1.9\times{10}^{18} $ years, so we can consider it as a stable nuclide in the s-process. We simplify the s-process path from $ ^{56}Fe $ to $ ^{183}W $ to a no-branching s-process. This is because on the s-process path, the abundance flow of the branching s-process network must converge to the main s-process path after passing through neutron capture and $ \beta $ decay. Therefore, when calculating the abundance of a certain stable nuclide, we can approximate it by the no-branching s-process (classical s-process), which is the main reason for the classical s-process analytical solution given by Clayton. D D et al. \cite{clayton1961neutron}. So, we use Clayton's method and give the expression of the abundance function  of $ ^{183}W $ similar to Ref.~\cite{pan2020the}. The expressions of $ \psi(^{183}W) $ are as follows:

\begin{equation} 
	m_{k}=\dfrac{(\sum_{i=1}^{k}\frac{1}{\sigma_{i}})^{2}}{\sum_{i=1}^{k}\frac{1}{\sigma_{i}^{2}}} , 
	\quad\quad
	\lambda_{k}=\dfrac{\sum_{i=1}^{k}\frac{1}{\sigma_{i}}}{\sum_{i=1}^{k}\frac{1}{\sigma_{i}^{2}}}. 
	\quad
\end{equation}

\begin{equation} 
	\psi(^{183}W)=\lambda \dfrac{(\lambda\tau)^{m-1}}{\Gamma(m)} e^{-\lambda\tau}. 
\end{equation}

Where, $ m_{k} $ and $ \lambda_{k} $ are determined by all Maxwellian-Averaged neutron capture cross sections involved in the s-process from $ ^{56}Fe $ to $ ^{183}W $. 

Similar to Ref.~\cite{ward1976s,wu2009analytical}, Jing Pan etal. \cite{pan2020the} gives the analytic solution of network system Eq.~(\ref{eq:4}) by constant variation method. Unlike the above references, we will give the numerical solution of Eq.~(\ref{eq:4}) in this paper. 

When solving equations by the direct integral with the constant variation method, a constant is generated in the denominator after the integration, which is the difference between the Maxwellian-Averaged neutron capture cross sections of the nuclides. Therefore, this presents an obstacle to the sensitivity analysis. 

Table \ref{tab:1} shows the Maxwellian-Averaged neutron capture cross sections of the nuclides involved in the reaction network \cite{dai1987chart,bao2000neutron}. We notice that there are nuclides with similar Maxwellian-Averaged neutron capture cross sections, such as $ ^{185}Re $ and $ ^{186}Re $, $ ^{186}Os $ and $ ^{188}Os $, $ ^{187}Re $ and $ ^{189}Os $, and so on. When performing sensitivity analysis, the neutron capture cross section data need to be adjusted up and down so that the neutron capture cross section data are extremely similar or even equal, and the difference between these cross sections is in the denominator of the expression for the analytical solution, which makes sensitivity analysis poses difficulties. For this reason, we solve the  Eq.~(\ref{eq:4}) with numerical methods.

\begin{table}[!h]
	\centering
	\caption{Maxwellian-Averaged neutron capture cross sections and half-life of each nuclide in the branching s-process.}\label{tab:1}%
	\begin{tabular}{@{}ccc@{}}\toprule
		nucleus &$<\sigma v>/v_{T} (mb)$ &half-life\\ \midrule
		$ ^{184}W $  & $ 224$    & stable\\
		$ ^{185}W $  & $ 703 $   & $75.1$ d \\
		$ ^{185}Re $ & $ 1535 $  &stable \\
		$ ^{186}Re $ & $ 1550 $  & $3.7186$ d \\
		$ ^{186}W $  & $ 176$    & stable\\
		$ ^{186}Os $ & $ 422$    & stable\\
		$ ^{187}Os $ & $ 896$    & stable\\
		$ ^{187}Re $ & $ 1160 $  & stable\\
		$ ^{188}Os $ & $ 399$    & stable\\
		$ ^{189}Os $ & $ 1168$   & stable\\  \bottomrule
	\end{tabular}
\end{table}

Usually, network equations are stiff equations \cite{hix2006thermonuclear}, so we use the fourth-order semi-implicit Runge-Kutta method \cite{yuan1987numerical} to calculate the numerical solutions in the following format.

\begin{equation}
	\left\{\begin{aligned}
		y_{n+1}=& y_{n}+\sum_{i=1}^{4} w_{i} K_{i}, \\
		K_{1}=& h\left[J\left(y_{n}\right)+b_{1} J\left(y_{n}\right) K_{1}\right], \\
		K_{2}=& h\left[f\left(y_{n}+\beta_{21} K_{1}\right)+b_{2} J\left(y_{n}+\eta_{21} K_{1}\right) K_{2}\right], \\
		K_{3}=& h\left[f\left(y_{n}+\beta_{31} K_{1}+\beta_{32} K_{2}\right)+b_{3} J\left(y_{n}\right.\right.+\eta_{31} K_{1}+\eta_{32} K_{2}) K_{3}], \\
		K_{4}=& h\left[f\left(y_{n}+\beta_{41} K_{1}+\beta_{42} K_{2}+\beta_{43} K_{3}\right)\right.\\
		&+b_{4} J\left(y_{n}+\eta_{41} K_{1}+\eta_{42} K_{2}+\eta_{43} K_{3}\right) K_{4}]. 
	\end{aligned}\right.
\end{equation}

Where $ J=\frac{\partial f}{\partial y} $ is the Jacobi matrix and $ h $ is the step. 

\begin{figure}[ht]
	\centering
	\includegraphics[scale=0.35]{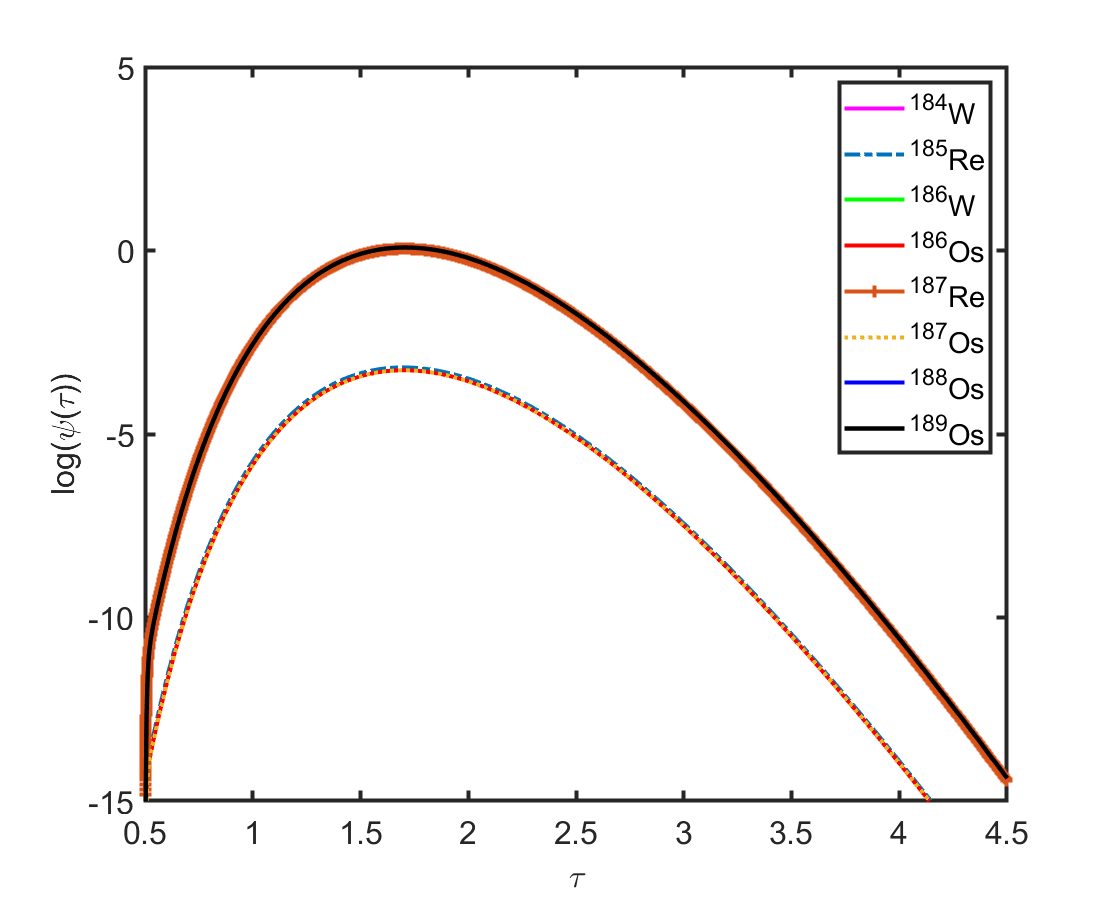}
	\caption{The abundance of each nuclide $ \psi(\tau)  $ versus $ \tau $ in the branching $ ^{187}Re $ -- $ ^{187}Os $ nuclear clock network.}
	\label{fig:2}
\end{figure}

In this way, we obtain the numerical solution of the network  Eq.~(\ref{eq:4}). The calculated results are shown in Figure \ref{fig:2}, where the values of the vertical coordinate abundances are taken as logarithms (Similar to Figures \ref{fig:3}  and \ref{fig:4}). 

\section{Sensitivity Analysis of Maxwellian-Averaged Neutron Capture Cross Section}\label{sec3}

Eight Maxwellian-Averaged neutron capture cross sections are involved in the simplified differential equations (Eq.~(\ref{eq:4})): $\sigma(^{184}W)$, $\sigma(^{185}Re)$, 
$\sigma(^{186}W)$, $\sigma(^{186}Os)$, $\sigma(^{187}Os)$, $\sigma(^{187}Re) $, $\sigma(^{188}Os) $, $\sigma(^{189}Os) $. At present, these data are measured by nuclear physics experimenters, but experiments usually have experimental errors. Therefore, we need to carry out sensitivity analysis on the Maxwellian-Averaged neutron capture cross section of the nucleus. It should be noted that the sensitivity analysis in this paper is indeed different from the sensitivity analysis in the traditional sense. Generally, sensitivity analysis is used to study and analyze the sensitivity of a system (or model) to changes in the state or output of the system parameters or conditions. Specifically, it is to change a parameter in the formula of the system (or model) and analyze the degree of change caused by the output of the system (or model), so as to judge the robustness of the system (or model) \cite{meerschaert2013mathematical}. However, we calculated the changes in the abundance of all nuclides after the neutron capture cross section parameters were changed, and used this change to identify the importance of nuclear reactions to the $ ^{187}Re $ -- $ ^{187}Os $ network. Therefore, in this paper, the sensitivity of a nuclear reaction cross section is defined as the sum of all the changes in nuclide abundance caused by the change of the cross section. 

We change the Maxwellian-Averaged neutron capture cross section of each nucleus from $ 80\% $ to $ 120\% $ in the step of $ H = 10\% $, and bring it into the original Eq.~(\ref{eq:4}) to get new results. Then the solution to the equation necessarily changes with each change in the neutron capture cross section data. We subtract the numerical solution obtained after the change from the original numerical solution without the change, and define the sensitivity as follows. 

\begin{equation} 
	D=\sum_{i=1}^{8}\sum_{j=1}^{4\times 10^{4}}|\psi_{ij}^{*}-\psi_{ij}|h .
	\label{eq:12}
\end{equation}

\begin{figure}[ht]
	\centering
	\includegraphics[scale=0.35]{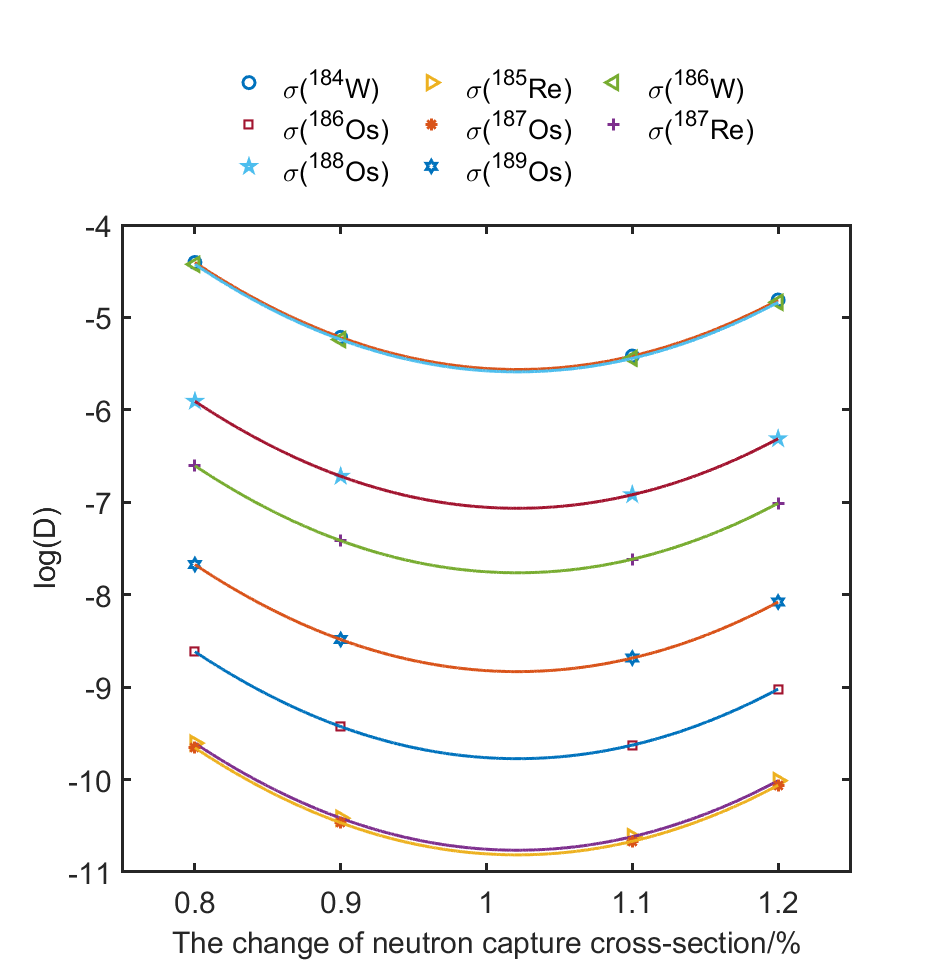}
	\caption{Sensitivity analysis of total reaction of branching network of $ ^{187}Re $ -- $ ^{187}Os $ nuclear clock.}
	\label{fig:3}
\end{figure}

Where $ \psi_{ij}^{*} $ is the numerical solution after changing the Maxwellian-Averaged neutron capture cross section and $ \psi_{ij} $ is the original numerical solution. The superscript of the first summation sign (inner summation sign) represents the number of computational nodes for the numerical solution of the equation. The products of the absolute values of the difference between the function values at all calculation nodes and the steps are summed, which is essentially the 1-norm of the two functions. In this paper, the value of $ \tau $ ranges from $ 0.5 $ to $ 4.5 $ in the calculation of integrals. The reason for starting from $ 0.5 $ is that the abundance values of all nuclides found below $ 0.5 $ are very small and have little contribution for integrals. The integral step is set as $ h=0.0001 $, so the number of nodes is $ 4 \times 10^{4} $. Eq.~(\ref{eq:12}) involves two summations. The innermost summation is the abundance change of a certain nuclide, and the outer summation is the sum of the abundance change of all the eight nuclides in the $ ^{187}Re $ -- $ ^{187}Os $ network. Therefore, Eq.~(\ref{eq:12}) represents the total abundance change of the whole $ ^{187}Re $ -- $ ^{187}Os $ network caused by changing the cross section parameters of a neutron capture reaction, which gives the total effect on the entire reaction network, as shown in Figure \ref{fig:3}.

Here, it should be pointed out that the number of nodes is determined by the step of the integral calculation. When the step is less than $ 0.001 $, the integral has converged. According to the numerical integration theory, the error of the integral is less than $ 10^{3} $. If higher accuracy is desired, the step should be smaller. In this paper, the step is $ h=0.0001 $. 

In this way, we can get the increasing order of Maxwellian-Averaged neutron capture cross sections which affect the total $ ^{187}Re $ -- $ ^{187}Os $ nuclear clock reaction network: $\sigma(^{184}W)$, $\sigma(^{186}W)$, $\sigma(^{188}Os)$, $\sigma(^{187}Re)$, $\sigma(^{189}Os)$, $\sigma(^{186}Os) $, $\sigma(^{185}Re) $, $\sigma(^{187}Os) $. Furthermore, the neutron capture reaction that has the greatest impact on the total $ ^{187}Re $ -- $ ^{187}Os $ nuclear clock reaction network is: $ {}^{184} W+n \rightarrow^{185} W $.

\begin{figure}[ht]
	\centering
	\includegraphics[scale=0.35]{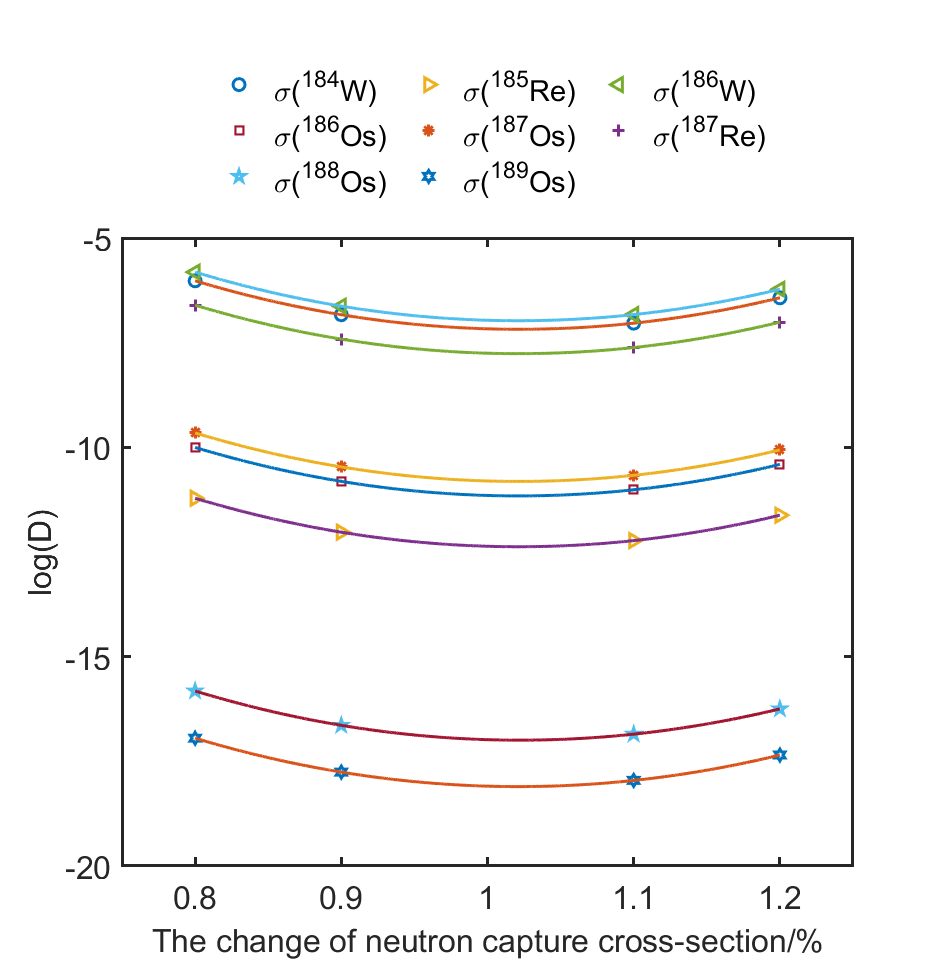}
	\caption{Sensitivity analysis of special nuclides ($ ^{187}Re $ and $ ^{187}Os $) in branchong network of $ ^{187}Re $ -- $ ^{187}Os $ nuclear clock. }
	\label{fig:4}
\end{figure}

In order to study the  $ ^{187}Re $ -- $ ^{187}Os $ nuclear clock problem better, we further analyze the sensitivity of the abundance of the special nuclides ($ ^{187}Re $ and $ ^{187}Os $) in the nuclear clock, and calculate Eq.~(\ref{eq:12}) again. However, it should be noted, that the superscript of the outer summation sign is 2, indicating that only the two nuclides ($ ^{187}Re $ and $ ^{187}Os $) are summed. The results are shown in Figure \ref{fig:4}.

Similarly, we obtain the increasing order of Maxwellian-Averaged neutron capture cross sections affecting the special nuclides ($ ^{187}Re $ and $ ^{187}Os $) is: $ \sigma(^{186}W) $, $ \sigma(^{184}W) $, $ \sigma(^{187}Re) $, $ \sigma(^{187}Os) $, $ \sigma(^{186}Os) $, $ \sigma(^{185}Re) $, $ \sigma(^{188}Os) $, $ \sigma(^{189}Os) $. Thus, the neutron capture reaction that has the greatest effect on the special nuclides ($ ^{187}Re $ and $ ^{187}Os $) is $ ^{186} W+n \rightarrow {}^{187}W $.

\section{Summary}\label{sec4}

The $ ^{187}Re $ -- $ ^{187}Os $ nuclear clock is an important cosmic nuclear clock for determining the age of the Galaxy. In this paper, we study the network equation calculation of the abundances of the nuclides associated with the $ ^{187}Re $ -- $ ^{187}Os $ nuclear clock in the s-process and the sensitivity analysis of the Maxwellian-Averaged neutron capture cross section for each nuclide. Due to the stiffness of the network equations, we give numerical solutions to the network equations by the semi-implicit Runge-Kutta method and perform detailed calculations. Our results are useful for the calibration of nuclear clock. 

In particular, using our numerical solutions and reaction flow of nuclides, this paper presents a detailed sensitivity analysis of the Maxwellian-Averaged neutron capture cross section for each nuclide of the $ ^{187}Re $ -- $ ^{187}Os $ nuclear clock reaction network in the s-process, and the results of the sensitivity analysis show that in the s-process, the neutron capture reaction with the greatest impact on the total branching reaction network of the $ ^{187}Re $ -- $ ^{187}Os $ nuclear clock is $ {}^{184} W+n \rightarrow {}^{185} W $, and the neutron capture reaction with the greatest influence on the two important nuclides $ ^{187}Re $ and $ ^{187}Os $ is $ ^{186} W+n \rightarrow {}^{187}W $. These results have positive significance for experimental nuclear physics, and we suggest that experimental nuclear physicists pay close attention to the measurements of these two Maxwellian-Averaged neutron capture cross sections.

\bibliography{mybibfile}

\end{document}